\documentclass[journal]{IEEEtran}
\usepackage{cite}
\usepackage{graphicx}
\usepackage{amsmath}
\usepackage{color}
\begin{document}
\newcommand{\cm}[1]{\ensuremath{ {{\rm cm}^{#1}}}}
\newcommand{\ntp}[2]{\ensuremath{#1\times10^{#2} } }
\newcommand{\asb}[2]{\ensuremath{#1_{\rm #2} }}

\begin{titlepage}

\thispagestyle{empty}
\def\thefootnote{\fnsymbol{footnote}}       

\begin{center}
\mbox{ }

\end{center}
\begin{center}
\vskip 1.0cm
{\Huge\bf
Comparison of Measurements of Charge
\vspace{2mm}

Transfer Inefficiencies in a CCD with 
}
\vspace{4mm}

{\Huge\bf
High-Speed Column Parallel Readout
}
\vskip 1cm
{\LARGE\bf 
Andr\'e~Sopczak$^1$,
Salim~Aoulmit$^2$,
Khaled Bekhouche$^1$,  
Chris~Bowdery$^1$,
Craig~Buttar$^3$,
Chris~Damerell$^4$, 
Dahmane~Djendaoui$^2$, 
Lakhdar~Dehimi$^2$, 
Rui~Gao$^6$, 
Tim~Greenshaw$^5$, 
Michal~Koziel$^1$,  
Dzmitry~Maneuski$^3$, 
Andrei~Nomerotski$^6$,
Nouredine~Sengouga$^2$,
Konstantin~Stefanov$^4$,
Tuomo~Tikkanen$^5$,
Tim~Woolliscroft$^5$, 
Steve~Worm$^4$,
Zhige~Zhang$^6$
\bigskip
}

\Large 
$^1$Lancaster University, UK\\
$^2$Biskra University, Algeria\\
$^3$Glasgow University, UK\\
$^4$STFC Rutherford Appleton Laboratory, UK\\
$^5$Liverpool University, UK \\
$^6$Oxford University, UK

\vskip 1.0cm
\centerline{\Large \bf Abstract}
\end{center}

\vskip 1.2cm
\hspace*{-0.5cm}
\begin{picture}(0.001,0.001)(0,0)
\put(,0){
\begin{minipage}{\textwidth}
\Large
\renewcommand{\baselinestretch} {1.2}
Charge Coupled Devices (CCDs) have been successfully used in several high energy physics 
experiments over the past two decades. Their high spatial resolution and thin sensitive 
layers make them an excellent tool for studying short-lived particles. The Linear Collider 
Flavour Identification (LCFI) Collaboration has been developing Column-Parallel CCDs for 
the vertex detector of a future Linear Collider which can be read out many times faster 
than standard CCDs. The most recent studies are of devices designed to reduce both the 
CCD's intergate capacitance and the clock voltages necessary to drive it. A comparative 
study of measured Charge Transfer Inefficiency values between our previous and new results
for a range of operating temperatures is presented.
\renewcommand{\baselinestretch} {1.}

\normalsize
\vspace{1.5cm}
\begin{center}
{\sl \large
Presented at the IEEE 2009 Nuclear Science Symposium, Orlando, USA, \\
and the 11th ICATPP Conference on Astroparticle, Particle, Space Physics, Detectors and \\
\vspace*{1mm}
Medical Physics Applications, Como, Italy, to be published in the proceedings.
\vspace{-6cm}
}
\end{center}
\end{minipage}
}
\end{picture}
\vfill

\end{titlepage}

\newpage
\thispagestyle{empty}
\mbox{ }
\newpage
\setcounter{page}{0}

\title{Comparison of Measurements of Charge Transfer Inefficiencies in a CCD with High-Speed Column Parallel Readout}
\author{Andr\'e~Sopczak,~\IEEEmembership{Member,~IEEE, }Salim~Aoulmit, Khaled~Bekhouche, 
Chris~Bowdery, Craig~Buttar, Chris~Damerell, Dahmane~Djendaoui, Lakhdar~Dehimi, Rui~Gao, 
Tim~Greenshaw, Michal~Koziel, Dzmitry~Maneuski, Andrei~Nomerotski, Nouredine~Sengouga, 
Konstantin~Stefanov, Tuomo~Tikkanen, Tim~Woolliscroft, Steve~Worm, Zhige~Zhang 
\thanks{A. Sopczak is with Lancaster University, UK. Presented on behalf of the}\thanks{~~LCFI Collaboration; 
E-mail: andre.sopczak@cern.ch}
\thanks{S. Aoulmit is with LMSM Laboratory Biskra University, Algeria}
\thanks{K. Bekhouche is with Lancaster University, UK}
\thanks{C. Bowdery is with Lancaster University, UK}
\thanks{C. Buttar is with Glasgow University, UK}
\thanks{C. Damerell is with STFC Rutherford Appleton Laboratory, UK}
\thanks{D. Djendaoui is with LMSM Laboratory Biskra University, Algeria}
\thanks{L. Dehimi is with LMSM Laboratory Biskra University, Algeria}
\thanks{R. Gao is with Oxford University, UK}
\thanks{T. Greenshaw is with Liverpool University, UK}
\thanks{M. Koziel is with Lancaster University, UK}
\thanks{D. Maneuski is with Glasgow University, UK}
\thanks{A. Nomerotski is with Oxford University, UK}
\thanks{N. Sengouga is with LMSM Laboratory Biskra University, Algeria}
\thanks{K. Stefanov is with STFC Rutherford Appleton Laboratory, UK}
\thanks{T. Tikkanen is with Liverpool University, UK}
\thanks{T. Woolliscroft is with Liverpool University, UK}
\thanks{S. Worm is with STFC Rutherford Appleton Laboratory, UK}
\thanks{Z. Zhang is with STFC Rutherford Appleton Laboratory, UK}}
\maketitle
\begin{abstract}
Charge Coupled Devices (CCDs) have been successfully used in several high energy physics 
experiments over the past two decades. Their high spatial resolution and thin sensitive 
layers make them an excellent tool for studying short-lived particles. The Linear Collider 
Flavour Identification (LCFI) Collaboration has been developing Column-Parallel CCDs for 
the vertex detector of a future Linear Collider which can be read out many times faster 
than standard CCDs. The most recent studies are of devices designed to reduce both the 
CCD's intergate capacitance and the clock voltages necessary to drive it. A comparative 
study of measured Charge Transfer Inefficiency values between our previous and new results
for a range of operating temperatures is presented.
\end{abstract}
\begin{IEEEkeywords}
LCFI, CPCCD, CCD, charge transfer inefficiency, radiation damage
\end{IEEEkeywords}
\section{Introduction}
The Nobel Prize-winning invention of an imaging semiconductor circuit 
(the CCD sensor)~\cite{nobel2009a,nobel2009b} has important applications for 
particle physics detectors. 
The study of radiation hardness is crucial for these 
applications~\cite{Damerell,Stefanov,LCFI_web}. The LCFI collaboration has 
been developing and testing new CCD detectors for about 10 years~\cite{Damerell,Stefanov,LCFI_web}. 
Previous experimental results on CCD radiation hardness were reported for example 
in~\cite{Marconi,Brau2000,Brau2004,Brau2005,Brau2007}. Several theoretical models have increased the 
understanding of radiation damage effects in CCDs~\cite{Mohsen,Hopkins,Hardy,Sopczak2008,Sopczak2009}. 
Simulation and modeling of CCD radiation hardness effects for a CCD prototype with sequential readout 
was reported at IEEE2005; comparing full TCAD simulations with analytic models was reported 
at IEEE2006; simulation and modeling of a CCD prototype with column parallel readout (CPCCD) 
was reported at IEEE2007 and in~\cite{Sopczak2008}. Experimental measurements using a 
method to determine the charge transfer inefficiency (CTI) were performed with a CPCCD 
prototype CPC-1 at a test stand at Liverpool University~\cite{Sopczak2009a}. This work 
focuses on a new CPCCD prototype, CPC-T, at a test stand at Oxford University. 
The high radiation environment near the interaction point at a future Linear Collider damages 
the CCD material which leads to defects acting as electron traps in the silicon. 
The radiation level at a Linear Collider is estimated to be $5\times10^{11}$ e/cm$^2$ 
and $10^{10}$ neutrons/cm$^2$ per year at the inner vertex 
detector layer (14 mm radius)~\cite{Maruyama,Vogel}.
The mechanism of creating traps has been discussed in the literature~\cite{Walker,Saks,Srour}.
These traps result in charge transfer inefficiency. 

The column parallel technology is in development to cope with the required readout rate. 
The CPC-T used is a 4-phase variant of the CPCCD technology capable of 50 MHz readout frequency. 
Experimental work at Liverpool University on an un-irradiated CPC-1 led to CTI values compatible 
with zero but with rather large uncertainties~\cite{Sopczak2009a}.
In this paper we demonstrate a method to determine the CTI value with an un-irradiated 
CPC-T aiming for small CTI uncertainties.

\section{Theory}
\begin{figure}
    \begin{center}
    \includegraphics[width=0.49\textwidth,origin=c,angle=0]{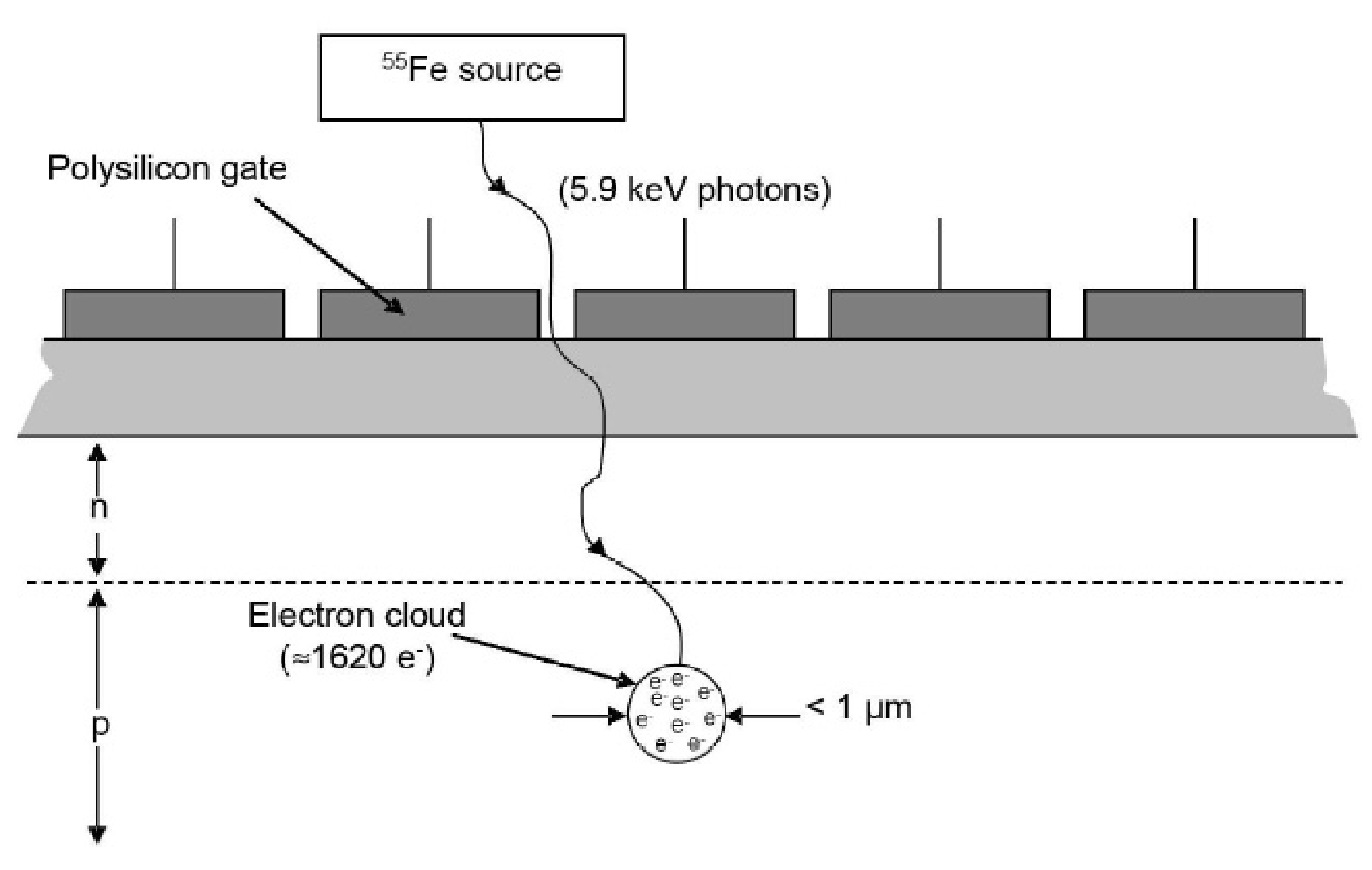}
    \end{center}
\vspace*{-4mm}
    \caption{$^{55}$Fe X-ray interacting with a CCD. A 5.9 keV photon generates an electron 
             cloud of approximately 1620~$e^{-}$.}
    \label{fig:XrayInteraction}
\end{figure}

{Soft X-ray photons (0.1 to 10 keV) interact with silicon atoms within the depleted layer. 
The depletion layer thickness is a parameter that determines the quantum efficiency at 
energies above 4 keV~\cite{Prigozhin}. The absorbed energy generates multiple e-h pairs. 
For a 5.9 keV X-ray source, one event (photon) generates a cloud of approximately 
1620 electrons (Fig.~\ref{fig:XrayInteraction}~\cite{Janesick}) 
contained within a diameter less than one micrometer~\cite{Tikkanen}. 
The charge from a single X-ray photon, 
generated within the depletion region of a target pixel, is not transferred completely to 
the next pixel due to two main effects: the generation of thermal dark charge within the 
depletion region and the trapping of signal charge within the n-buried channel~\cite{Hopkinson}. 
Since the buried channel is within the depletion layer, the important mechanisms 
are the capture of signal from the conduction band to the trap level and their 
subsequent emission back to the conduction band~\cite{Hopkins94}.  Therefore, the X-ray event exhibits 
a `tail' of deferred charge. Also, the charge generated in the field-free region diffuses 
into neighboring pixels and adds to the `tail' of deferred charge. The size and shape of 
this tail is a sensitive indicator of charge transfer inefficiency. X-ray stimulation is 
therefore extremely valuable in characterizing the CTI~\cite{Janesick}.
Many analyses have been made to simulate the effect of traps via the emission and capture 
processes~\cite{Hopkinson, Hardy,Kono}. The following simplified equations, 
based on earlier work by Shockley, Read and Hall~\cite{SR,Hall}, have been used to
analyse the CTI:}
\begin{equation*}
\frac{dn_t}{dt}=-\frac{n_t}{\tau_e}+\frac{(N_t-n_t)}{\tau_c}
\end{equation*}
\begin{equation*}
\tau_c=\frac{1}{\sigma_nv_{th}n_e}
\end{equation*}
\begin{equation*}
\tau_e=\frac{\exp(E_t / kT)}{\sigma_nv_{th}N_C}
\end{equation*}
where $n_t$ is the density of filled traps, $N_t$ is the total density of traps, 
$E_t$ is the trap energy level below the bottom of the conduction band, $\tau_e$ is the 
emission time constant, $\tau_c$ is the capture time constant,
 $\sigma_n$ is the trapping cross section, $v_{th}$ is the thermal velocity of carriers,
 $N_C$ is the effective density of states in the conduction band and $n_e$ is the density of electrons in the 
conduction band. For a detailed analytic model, 
the following parameters have been taken into account:
\begin{itemize}
\item the order of magnitude of the emission and capture time constants compared to the shift 
      time (time needed for a charge packet to move from one pixel to another). 
\item the shape of the electrostatic potential, which can be assumed to be placed in the 
      middle of the well.
\item the level of the signal charge (density of free electrons) within the 
      potential well in comparison to the total density of traps.
\end{itemize}

\section{Test Stand for CCD Operation}
\begin{figure}[bhp]
    \begin{center}
    \includegraphics[width=0.49\textwidth,height=8.5cm,origin=c,angle=0]{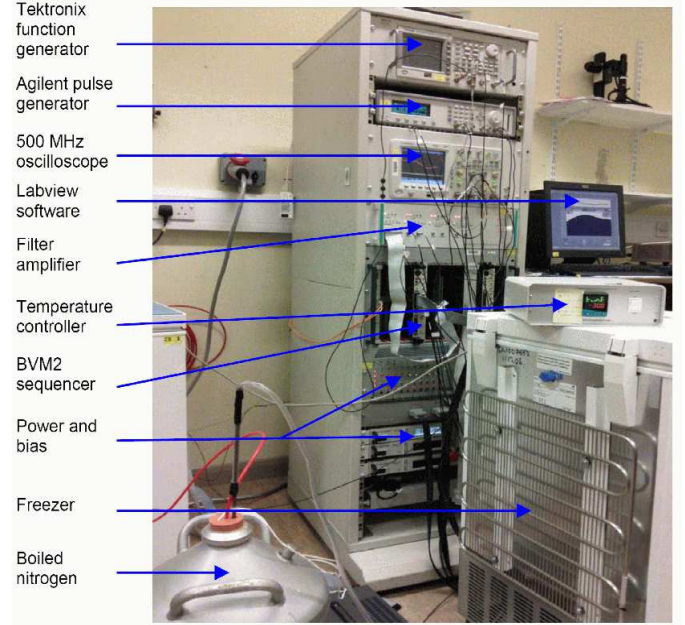}
    \end{center}
    \caption{Picture of the CPC-T readout. The CPC-T mother board is inside the freezer.}
    \label{fig:CPCTpicture}
\end{figure}

\begin{figure}
    \begin{center}
    \includegraphics[width=0.49\textwidth,origin=c,angle=0]{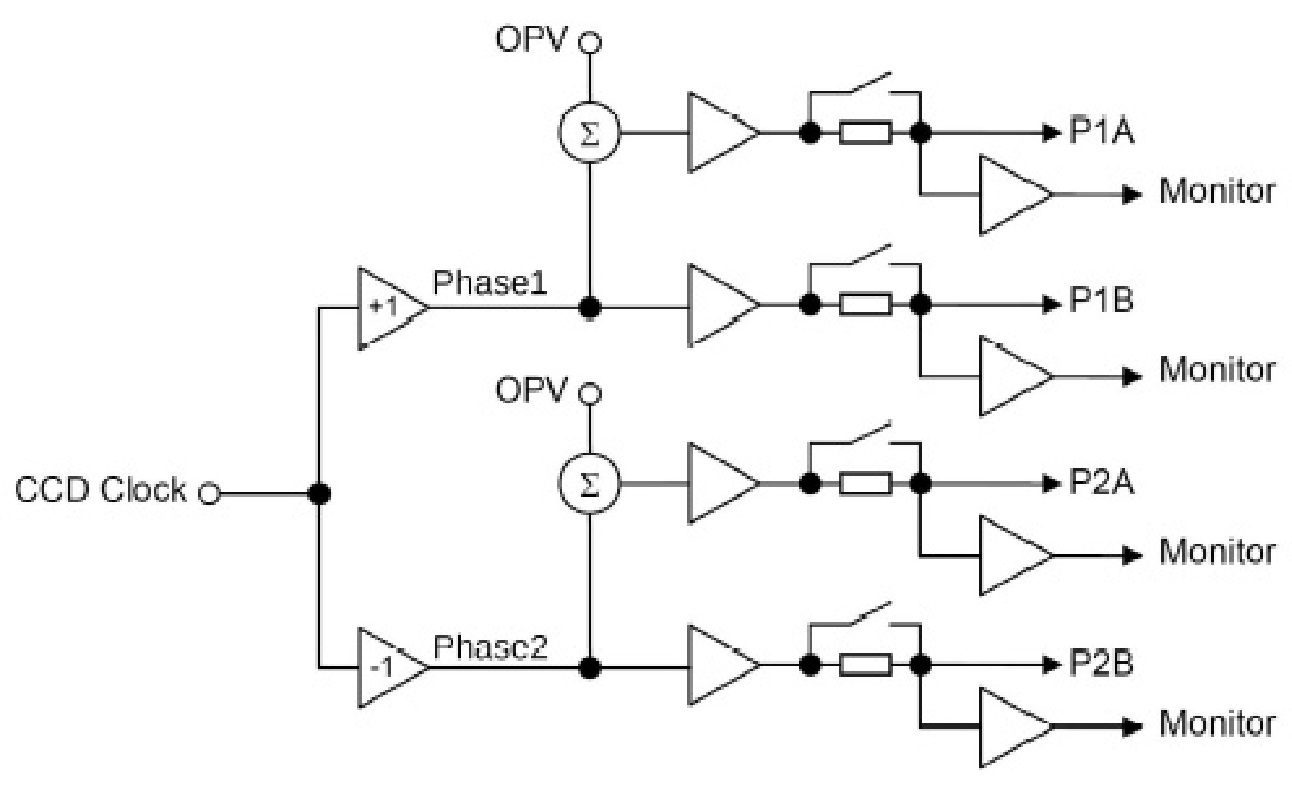}
    \end{center}
    \caption{Clocking scheme for the 4-phase variant. This variant is driven as a two-phase CCD
             with additional DC voltage (OPV) to the voltage of the first gate of a pixel.}
    \label{fig:CPCTvariant}
\end{figure}

\begin{figure}
\vspace*{5mm}
    \begin{center}
    \includegraphics[width=0.49\textwidth,origin=c,angle=0]{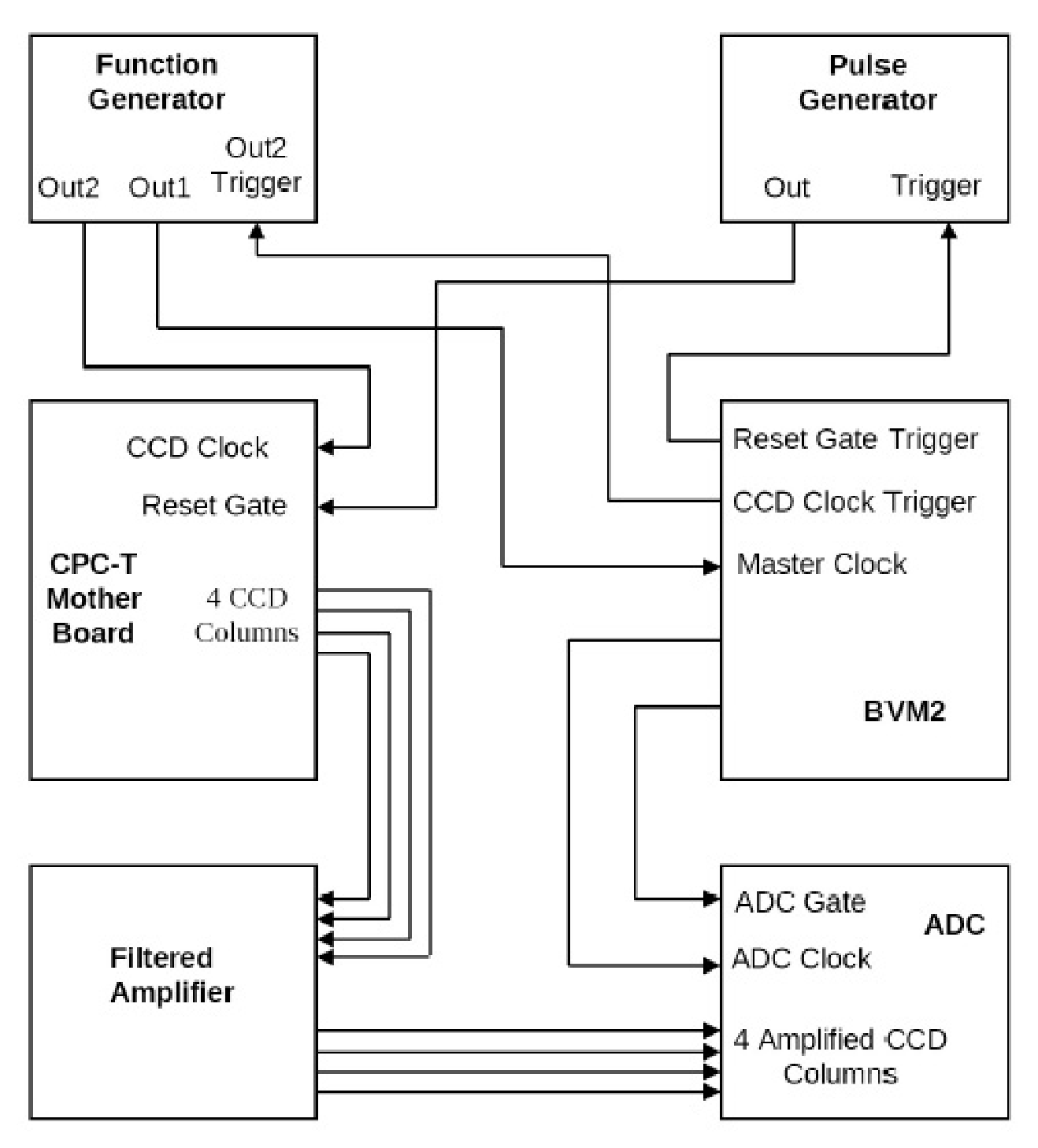}
    \end{center}
    \caption{Schematic diagram of the CPC-T readout.} 
    \label{fig:SchematicDiagram}
\vspace*{5mm}
\end{figure}

A test stand has been set up with readout electronics and a freezer unit {as shown in Fig.~\ref{fig:CPCTpicture}}. 
The temperature range of the freezer is from room temperature down to about $-60~^\circ$C. 
Fine control of the CPC-T temperature is done using a CAL9500P controller (the temperature is kept constant 
within $0.1~^\circ$C). A flux of boiled nitrogen is introduced into the motherboard box 
to purge water vapour. {The CPC-T chips come in 2 main variants: inherent 4-phase CCD driven as 2-phase CCD, 
and `pedestal' 2-phase CCD with 2 additional DC-biased gates. The former was used for this measurement. 
The first and second gates of each pixel, P1A and P1B, are driven by Phase1,
and P1A is offset by the DC voltage OPV (Offset and Pedestal Voltage), 
as shown in Fig.~\ref{fig:CPCTvariant}}.
The CPC-T has $500\times10$ pixels with a pixel size of $20\times 20~\mu$m$^2$. 
Initial measurements have been performed on an un-irradiated device in standalone mode, 
where the signals from four columns of the CCD were amplified and connected to 
external 14-bit ADCs. A $^{55}$Fe source emitting 5.9~keV X-rays was attached to a 
holder at a distance of 1~cm from the CCD to provide the signal charge. The schematic 
diagram in Fig.~\ref{fig:SchematicDiagram} illustrates the electronics used to drive 
and read out the CPC-T. The apparatus is controlled by a LabView program through 
interface modules. The BVM2 sequencer receives the master clock of 1~MHz from the function generator 
to provide four signals, two for ADC and two to trigger the generators which produces 
a CCD clock and reset gate signals. The 2-resets configuration, when one reset is applied 
before reading the first pixel of the CCD and one after reading the last pixel, is used 
in this measurement. This configuration leads to low noise since the reset noise 
is absent. 
The occupancy is about 1\% for the integration time of 100~ms given the strength 
of the X-ray source and the experimental layout.
The number of frames (complete readout of the CCD) is not kept constant to study the effect 
on the statistical uncertainty. Our method is based on the typical methods used for 
serially read out CCDs, where the CTI is determined by fitting a line to the readout 
charge signal as a function of the pixel number.
A linear function can be expected when the CTI is small.

\section{Signal Measurement and CTI Determination Method}
\begin{figure}[hbp]
\vspace*{1cm}
    \begin{center}
    \includegraphics[width=0.49\textwidth,origin=c,angle=0]{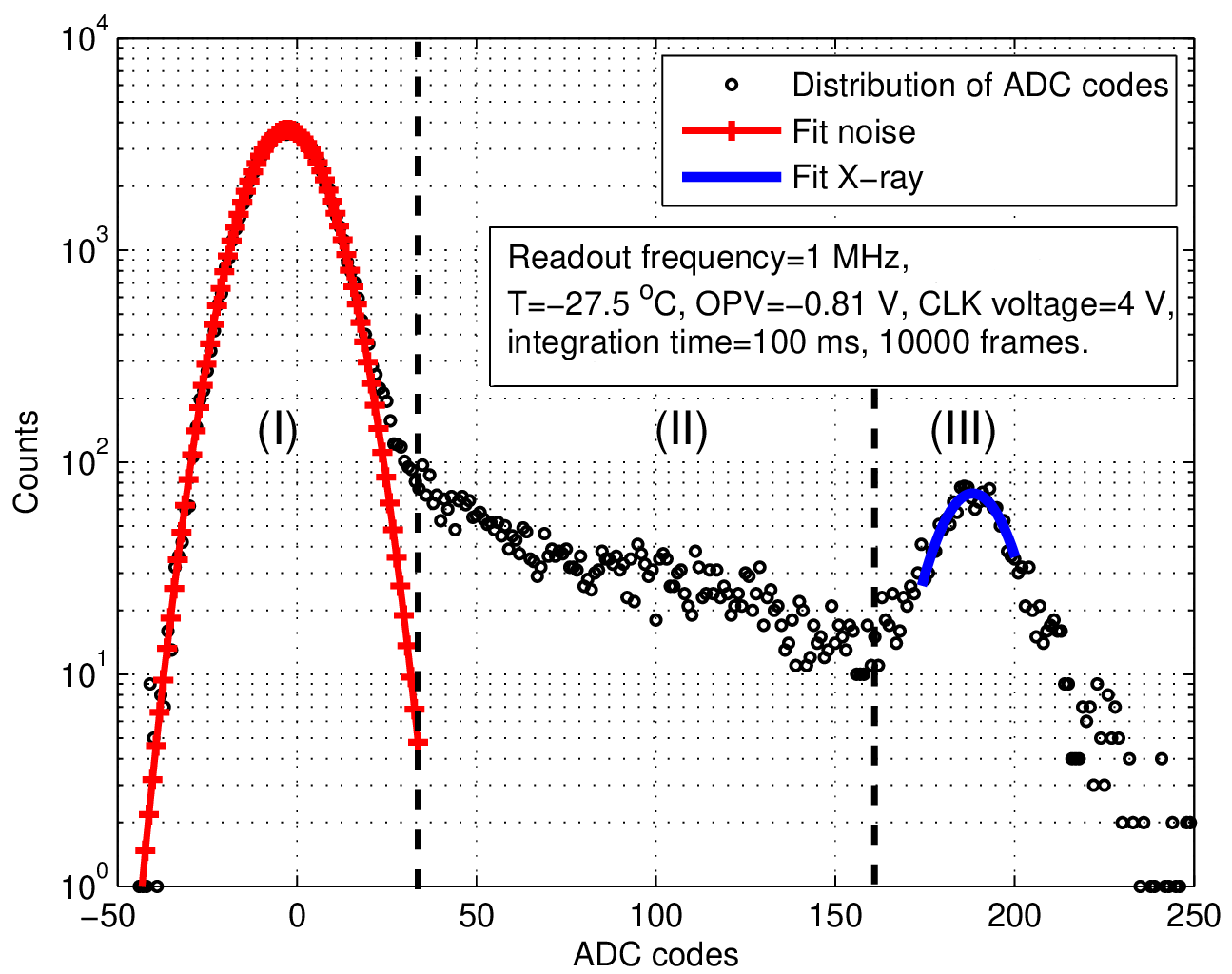} \hfill
    \end{center}
\caption{\label{fig:method}
Distribution of ADC codes for channel (column) two. Three regions are observed: 
(I) the high peak region which represents the noise, 
(II) the region separating the two peaks which represents the charge sharing between pixels, and 
(III) the X-ray peak region which represents the fully collected charge in a single pixel. 
The noise and X-ray peaks are fitted by a Gaussian function to determine the centroids. 
The X-ray signal is the difference between the two centroids.}
\vspace*{0.5cm}
\end{figure}

\begin{figure}[tp]
    \begin{center}
    \includegraphics[width=0.49\textwidth,origin=c,angle=0]{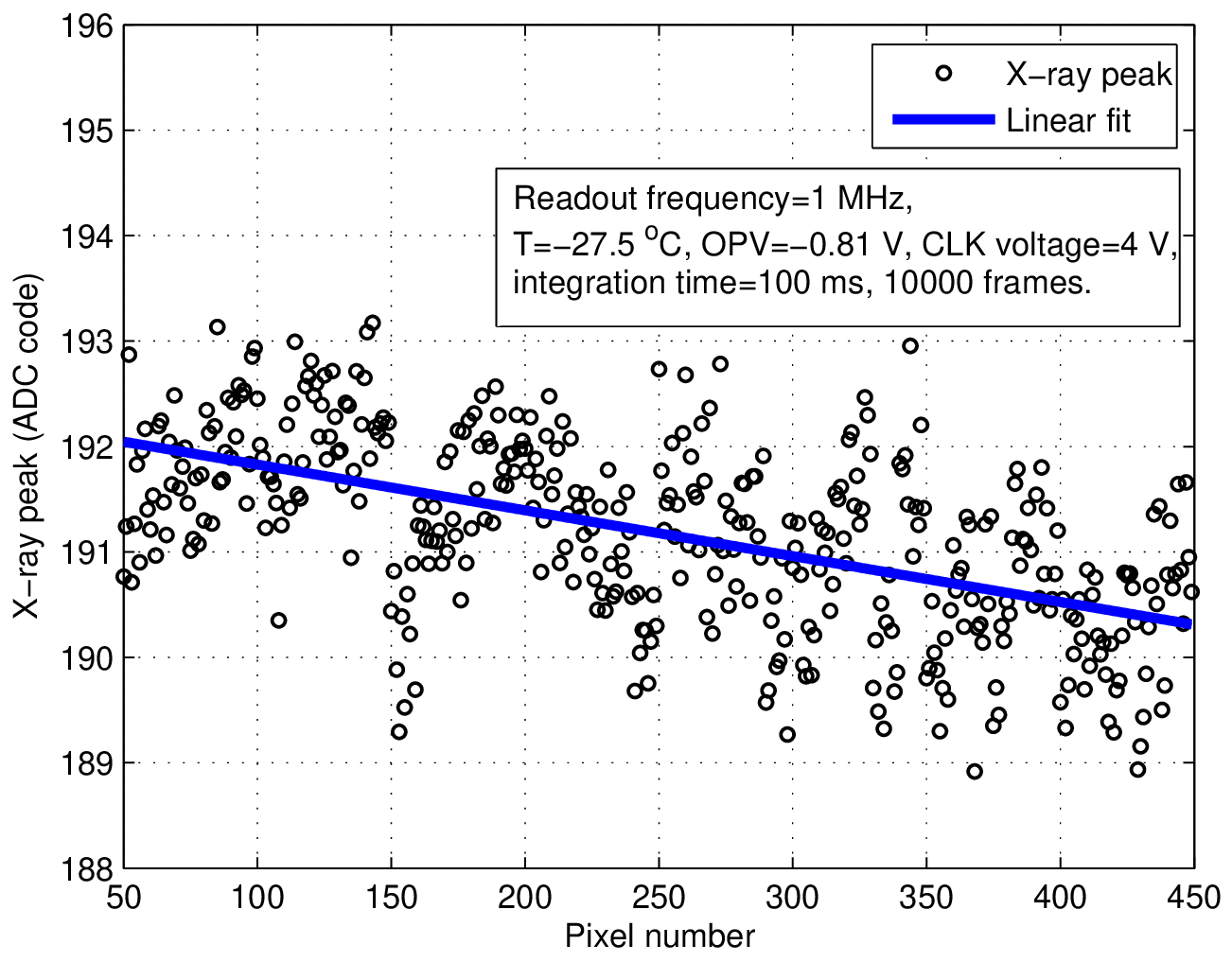}
    \end{center}
\caption{\label{fig:method2}
Linear fit to average ADC codes. The X-ray centroid of each pixel is calculated by averaging 
ADC codes within the interval $x_0-n\sigma$ and $x_0+n\sigma$ where 1$\leq$n$\leq$3.}
\end{figure}

The fast ADCs convert the signal charge after amplification with a wideband preamplifier. 
The four columns are read out in 4 channels by ADCs. The signal charges of 500 pixels were 
acquired in 1000 and 10000 frames per measurement for each temperature. 
Using 10000 frames leads to a sufficient 
statistical precision (around few times $10^{-6}$ for the CTI measurement). The collected data have been
analysed using MATLAB~\cite{matlab}. First, we begin by applying correlated double sampling, where 
the difference between the signals of two consecutive pixels is taken to be the signal 
charge collected by the latter pixel. 
This reduces some of the noise components (e.g. 1/f, kTC, white noise, etc.) 
in the CCD signal. 
As an example, 
Fig.~\ref{fig:method} shows the pulse-height distribution of ADC codes for the last 10 adjacent 
pixels in column 2. The charge transfer inefficiency (CTI) 
in one pixel is defined as the ratio of signal lost during transfer (captured by traps) 
to the initial signal charge.
The CTI is calculated following these steps: 
creation of a histogram with ADC codes of 10 pixels in a column. 
These pixels have nearly the same baseline.
The histogram creation is repeated 50 times to cover all 500 pixels in a column. 
Fits with Gaussian functions are made to the noise and X-ray peaks.
We use $x_0-n\sigma$ and $x_0+n\sigma$ as limits for noise peak and X-ray peak 
($x_0$ and $\sigma$ are the centroid and the standard deviation respectively of 
the two Gaussian functions resulting from fitting the noise and X-ray peaks)
to determine noise centroid and X-ray centroid for each pixel. The factor $n$ is 
chosen between 1 and 3 depending on the amount of charge sharing.
The X-ray signal for each pixel is the difference between the X-ray peak 
centroid and the noise peak centroid. 
Figure~\ref{fig:method2} shows the distribution of the averages as a function of pixel number. 
The distribution is fitted with the first-order 
polynomial function $P_0+P_1j$, where $P_0$ corresponds to the charge at the first pixel, 
$P_1$ is the slope and $j$ is the pixel number. 
The CTI is determined using $CTI=-P_1/P_0$.

\section{CTI Results Pre-Irradiation}
\begin{figure}
    \begin{center}
    \includegraphics[width=0.49\textwidth,origin=c,angle=0]{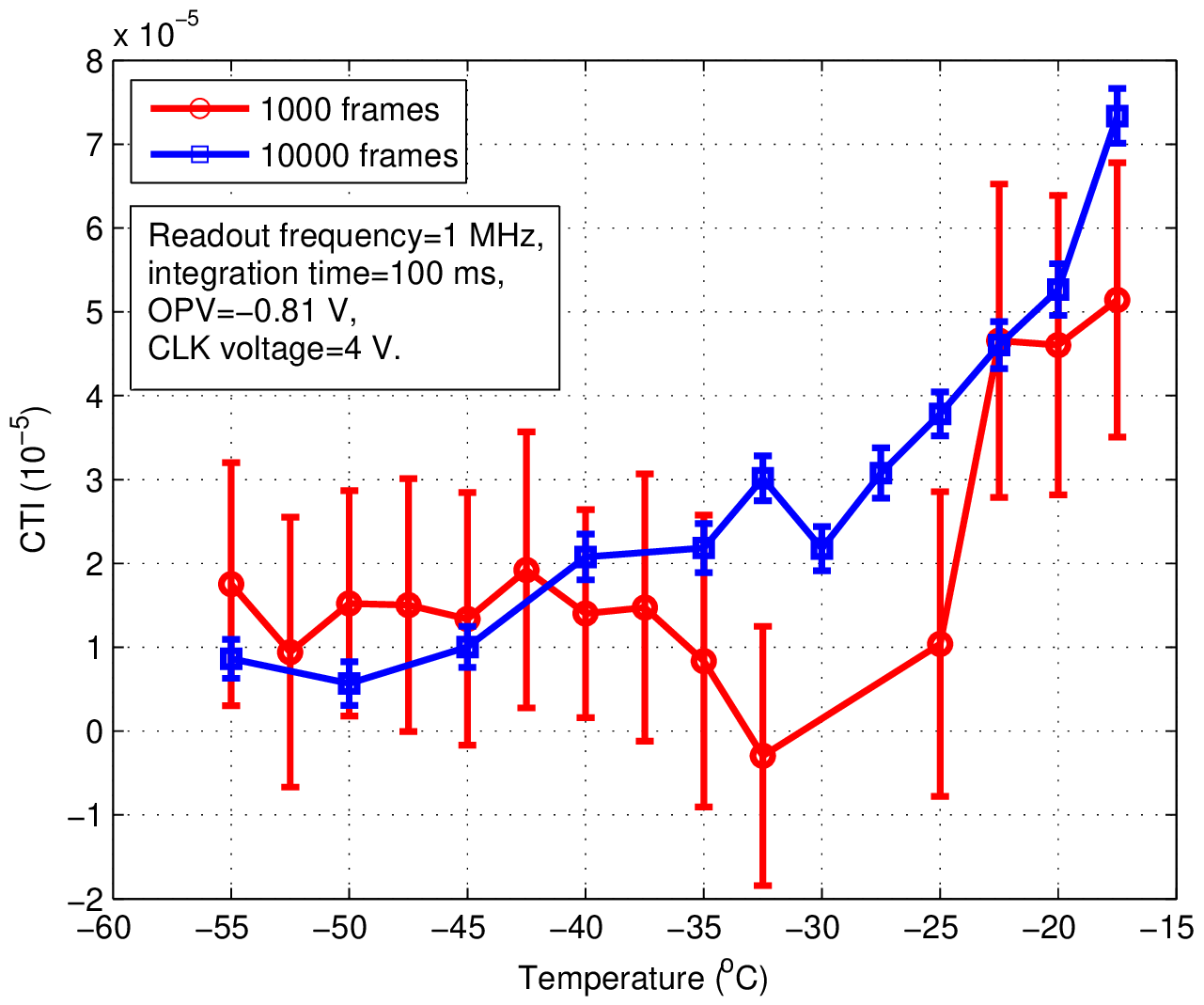} \hfill
    \end{center}
\caption{\label{fig:results}
CTI as a function of temperature for different numbers of frames. 
The error bars have been significantly reduced by increasing the number of frames.}
\end{figure}

\begin{figure}
    \begin{center}
    \includegraphics[width=0.49\textwidth,origin=c,angle=0]{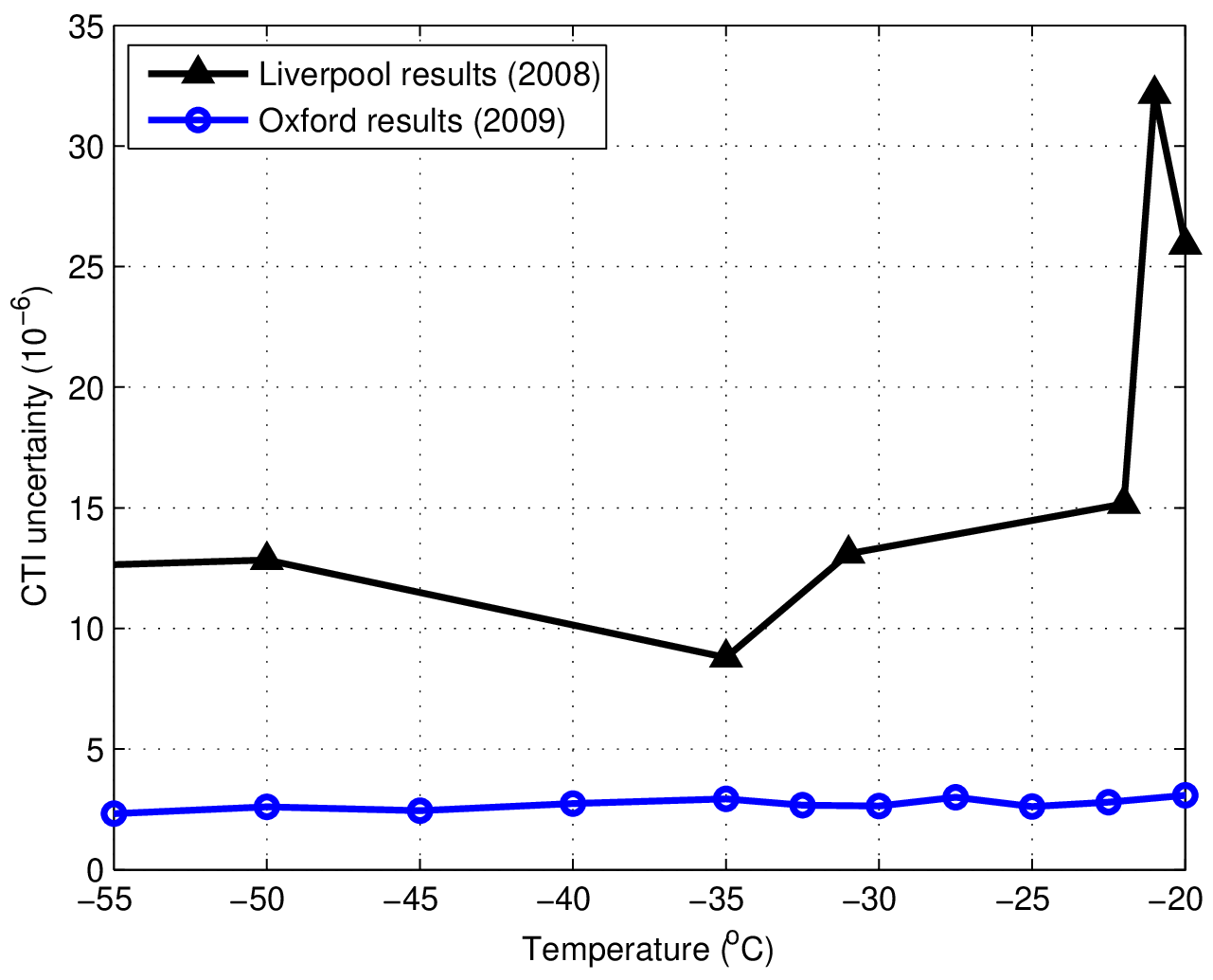}
    \end{center}
\caption{\label{fig:results2}
Comparison of the Oxford results with 10000 frames with the Liverpool 
results~\cite{Sopczak2009a} where the number of frames was 5000 and a fraction of 
data was lost because of a sampling inefficiency.}
\end{figure}

\begin{figure}
    \begin{center}
    \includegraphics[width=0.49\textwidth,origin=c,angle=0]{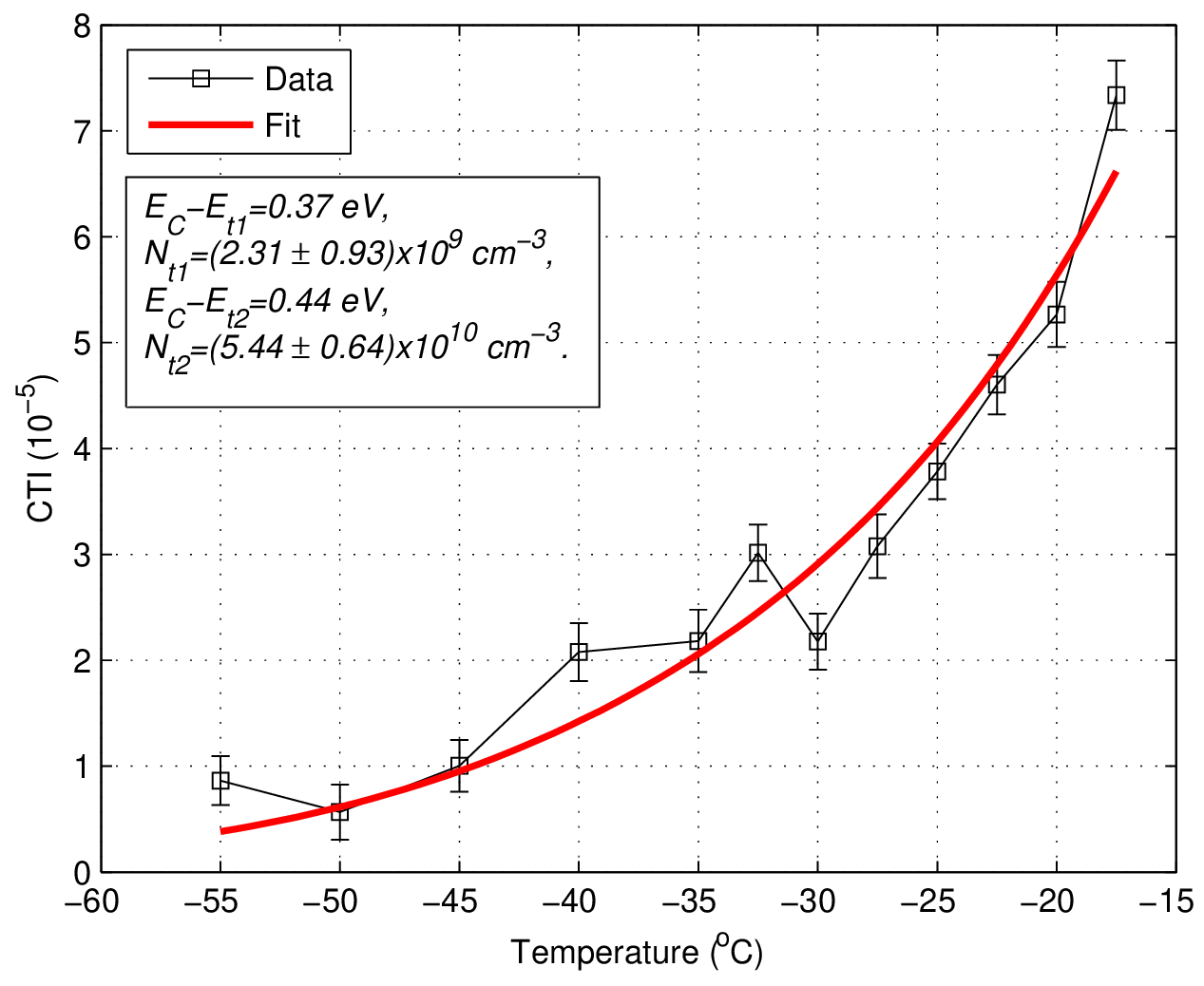}
    \end{center}
    \caption{Non-linear fit of the measured CTI using our analytic model~\cite{Sopczak2009}.
             The model includes two acceptor traps, 0.37 and 0.44~eV below the conduction band. 
	     A trapping cross-section
             $\sigma_n=3~\times~10^{-15}$~cm$^2$ is used for both traps.} 
    \label{fig:CTI2LevelFit}
\end{figure}

Figure~\ref{fig:results} shows the CTI values as a function of temperature for an
 un-irradiated CPC-T for 1000 and 10000 frames. The CTI has been calculated using 
a linear fit of average ADC codes versus pixel number. Figure~\ref{fig:results2} shows a 
comparison with the CPC-1 measurement~\cite{Sopczak2009a} taken at the test stand in Liverpool. 
Uncertainties have been reduced mostly by increasing the number of frames. For this 
CCD with 500 pixels per column a CTI value of $10^{-5}$ means that only $0.5\%$ of the signal charge is lost, 
which is acceptable in normal operation. 
{The apparent trend of the CTI at high temperatures in the operating range used is probably due 
to the contribution of two effects. 
First, there is the effect of thermal carrier generation (dark current) which is highly temperature-dependent. 
The dark current, non-uniform by nature, can have a large effect on the signal charge transfer for 
high temperatures, long integration time and large number of pixels in the column~\cite{Albert}. 
Second, there is the possibility of the presence of low trap density 
that could have been created during the long duration (around two years) of exposure to a soft X-ray 
source while studying the device. This significant positive value of CTI before irradiating 
the CCD was observed experimentally and modeled by a simple analytic 
model by including one trap level~\cite{Stefanov}.
Using our analytic model~\cite{Sopczak2009}, the CTI is expressed as}
\begin{eqnarray*}
CTI&=&2\frac{N_t}{n_s}[1-\exp(-t(\frac{1}{\tau_c}+\frac{2}{\tau_e}))]\times\nonumber\\
& &[(\frac{\tau_s}{\tau_e}\frac{(1-\exp(-\frac{t}{\tau_s}))}{(1-\exp(-t(\frac{1}{\tau_s}+\frac{1}{\tau_e})))})\exp(-\frac{t}{\tau_e})\nonumber\\
& &-\exp(-\frac{t_w}{\tau_e})].
\end{eqnarray*}
{We have fitted the CTI curve including two deep traps as shown in~Fig.~\ref{fig:CTI2LevelFit}. 
Both are electron traps at 0.37~eV and 0.44~eV below the bottom of the conduction 
band and having a trapping 
cross-section $\sigma_n=3\times10^{-15}$~cm$^2$. We have considered
that there is no interaction between the two traps, so they interact independently with the
signal charge. Therefore, the total CTI is the sum of CTIs resulting from the effect of the traps.
The fit is in good agreement with the data and shows that the 0.44~eV trap is the dominant one 
in this range of temperature as its density ($N_{t2}=5.22\times10^{10}$~cm$^{-3}$) is much larger 
than that of the 0.37~eV trap ($N_{t1}=2.63\times10^{9}$~cm$^{-3}$).}

\begin{figure}
    \begin{center}
    \includegraphics[width=0.49\textwidth,origin=c,angle=0]{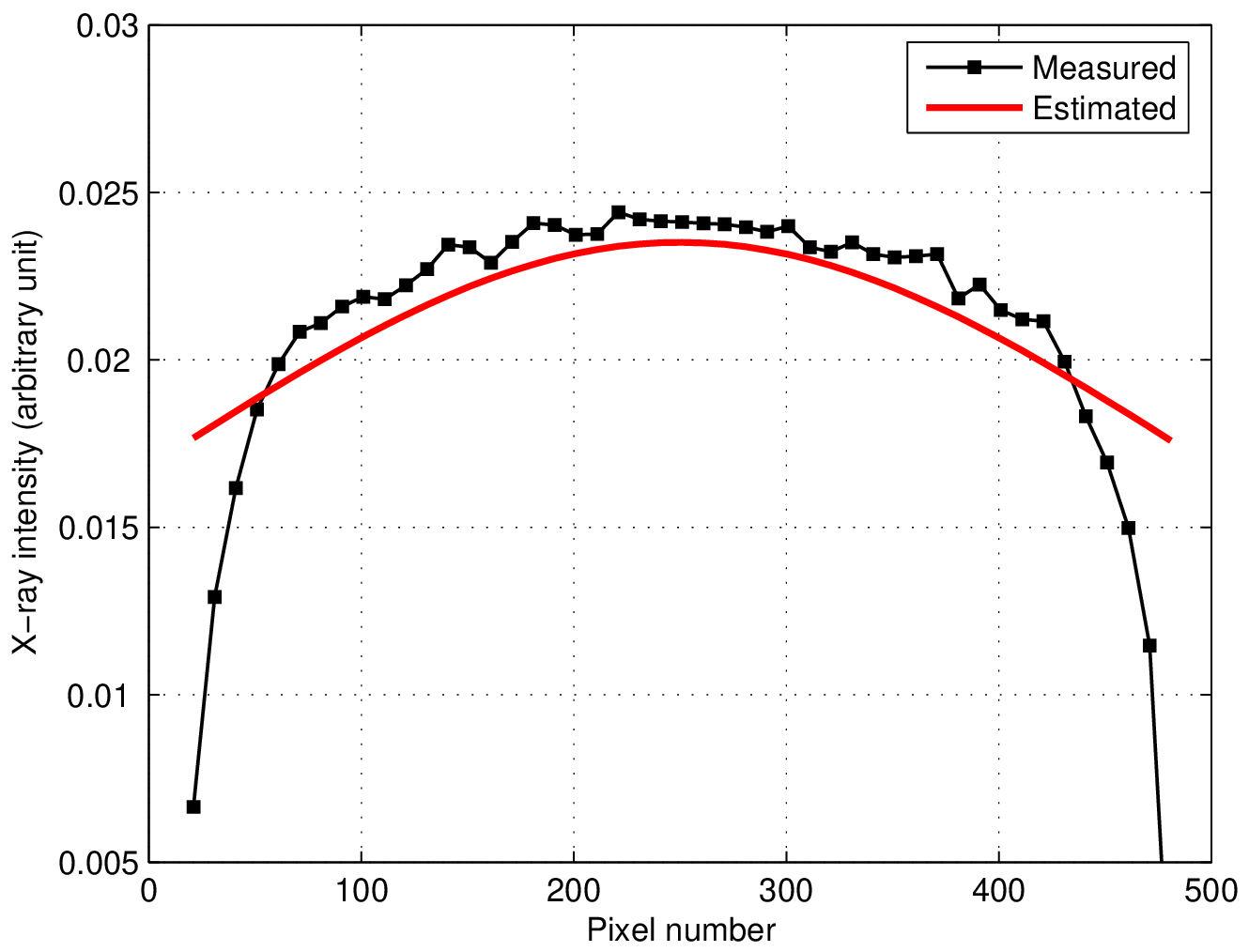}
    \end{center}
    \caption{X-ray distribution upon CPC-T. 
Measured distribution is determined by counting all ADC codes above the 
noise threshold which represent the energy deposited by X-rays. Simple formula reflecting the geometry 
is used to estimate the distribution.}
    \label{fig:SourceUniformity}
\end{figure}

The accuracy of the CTI calculation can be improved by positioning the $^{55}$Fe source so 
that it irradiates uniformly the CCD, carefully choosing the gain to use the ADC in its maximum 
dynamic range and acquiring data in a large number of frames.
The non-uniformity of the X-ray source coverage 
has an effect on the CTI determination. This is well understood and reproduces the estimate 
based on a geometry where the source is placed 1~cm away from the CPC-T. 
{Figure~\ref{fig:SourceUniformity} shows the non-uniformity of the X-ray source coverage. The 
figure contains two curves, the measured and the estimated X-ray distributions. The measured 
distribution is determined by counting all ADC codes above the noise threshold $x_0+3\sigma$. 
The estimated distribution is determined using the following formula for the given geometry:}
\begin{equation*}
f(n)=\frac{\bigl(h/l_p\bigr)^2}{\biggl(\bigl(h/l_p\bigr)^2+\bigl(n-n_0\bigr)^2\biggr)^{3/2}}
\end{equation*}
{where $f$ is the distribution of X-rays upon the CPC-T, $n$ is the pixel number, $h$ is the 
distance between source and CPC-T, $l_p$ is the length of one pixel and $n_0$ is the pixel 
number corresponding to the vertex position. In order to avoid the effect of the non-uniformity, 
the first and last 50 pixels are excluded from the fit to the averages 
for the CTI  determination (Fig.~\ref{fig:method}).}

\vspace*{0.5cm}
\section{Conclusions and Outlook}
An un-irradiated CPC-T was operated in a range of temperatures from $-15~^\circ$C to $-60~^\circ$C 
(freezer cooling) with different numbers of frames, 1000 and 10000. The CTI is analysed 
at different operating temperatures. A clear X-ray signal is extracted by calculating 
the difference between the noise centroid (baseline) and the X-ray centroid after well 
determining them. The statistical uncertainties have been reduced compared to a previous work 
with CPC-1~\cite{Sopczak2009a}.
The reduced uncertainties are due to the improved method, as we do not expect
a different behavior between CPC-1 and CPC-T regarding the CTI measurements as both devices have
a very similar geometric structure.
{The measurement of significant non-zero CTI values are indicators of trapping and 
thermally generated electrons. The former is dominant in this range of low temperatures. 
We expect that the CTI will increase after irradiation as the trap density increases.
The non-uniformity of the X-ray radiation has to be taken in account when measuring the CTI
of an irradiated CPC-T in the future.}

\section*{Acknowledgment}
We would like to thank Alex Chilingarov and Alex Finch for discussions and comments on the manuscript.
This work is supported by the Science and Technology Facilities Council (STFC) and 
Lancaster University. We would like to thank Oxford University for their hospitality.


\begin{thebibliography}{40}
\bibitem{nobel2009a} W.S.~Boyle and G.E.~Smith, ``Charge Coupled Semiconductor Devices'',
Bell Systems Technical Journal 49 (1970) 587. 
\bibitem{nobel2009b} G.F.~Amelio, M.F.~Tompsett and G.E.~Smith, ``Experimental Verification 
of the Charge Coupled Concept'', Bell Systems Technical Journal 49 (1970) 593.
\bibitem{Damerell} C.J.S. Damerell, ``Radiation damage in CCDs used as particle 
detectors'', ICFA Instrum. Bull. 14 (1997) 1.
\bibitem{Stefanov} K. Stefanov, PhD thesis, Saga University (Japan),
``Radiation damage effects in CCD sensors for tracking applications
in high energy physics'', 2001.
\bibitem{LCFI_web} LCFI collaboration homepage: http://hepwww.rl.ac.uk/lcfi/.
\bibitem{Marconi} M.S. Robbins ``The Radiation Damage Performance of Marconi CCDs'', Marconi Technical Note S\&C 906/424 2000 (unpublished).
\bibitem{Brau2000} J.E. Brau and N.B. Sinev, ``Operation of a CCD particle detector in 
the presence of bulk neutron damage'', IEEE Trans. Nucl. Sci. 47 (2000) 1898.
\bibitem{Brau2004} J.E.~Brau, O.~Igonkina, C.T.~Potter, N.B.~Sinev, ``Investigation of 
radiation damage in the SLD CCD vertex detector'', IEEE Trans. Nucl. Sci. 51 (2004) 1742.
\bibitem{Brau2005}J.E.~Brau, O.~Igonkina, C.T.~Potter and N.B.~Sinev, ``Investigation 
of radiation damage effects in neutron irradiated CCD'', Nucl. Instr. and Meth. A549 (2005) 117.
\bibitem{Brau2007} J.E.~Brau, O.~Igonkina, N.B.~Sinev, J.~Strube, ``Investigations 
into properties of charge traps created in CCDs by neutron and electron irradiation'', 
Pramana-J. Phys. 69 (2007) 1093.
\bibitem{Mohsen} A.M. Mohsen and M.F. Tompsett, ``The effect of bulk traps on the performance 
of bulk channel charge-coupled devices'', IEEE Trans. Electron Dev. ED21, 11 (1974) 701.
\bibitem{Hopkins} I.H. Hopkins, G. Hopkinson and B. Johlander, ``Proton-induced charge transfer degradation in 
CCD's for near-room temperature applications'', IEEE Trans. Nucl. Sci. 41 (1994) 1984.
\bibitem{Hardy} T. Hardy, R. Murowinski and M.J. Deen, ``Charge transfer efficiency 
in proton damaged CCD's'', IEEE Trans. Nucl. Sci. 45 (1998) 154.
\bibitem{Sopczak2008} A. Sopczak et al., ``Radiation Hardness Studies in a CCD with 
High-speed Column Parallel Readout'', JINST 3 (2008) 5007.
\bibitem{Sopczak2009} A. Sopczak et al., ``Modeling of Charge Transfer Inefficiency 
in a CCD with High-Speed Column Parallel Readout'', IEEE Trans. Nucl. Sci. 56 (2009) 1613.
\bibitem{matlab} MATLAB http://www.mathworks.com 
\bibitem{Sopczak2009a} A. Sopczak et al., ``Measurements of Charge Transfer 
Inefficiency in a CCD with High-Speed Column Parallel Readout'', IEEE Trans. Nucl. Sci. 56 (2009) 2925.
\bibitem{Maruyama} T. Maruyama, Private Communication, Stanford Linear Accelerator Center (SLAC) 2006.
\bibitem{Vogel} A. Vogel, Private Communication (DESY Hamburg) 2006.
\bibitem{Walker} J.W. Walker and C.T. Sah, ``Properties of 1.0-MeV-electron-irradiated defect centers in silicon'', Phys. Rev. B7 (1972) 4587.
\bibitem{Saks} N.S.~Saks, ``Investigation of Bulk Electron Traps Created by Fast
 Neutron Irradiation in a Buried N-Channel CCD'', IEEE Trans. Nucl. Sci. 24 (1977) 2153.
\newpage
\bibitem{Srour} J.R.~Srour, R.A.~Hartmann and S.~Othmer, ``Transient and Permanent 
Effects of Neutron Bombardment on a Commercially Available N-Buried-Channel CCD'', 
IEEE Trans. Nucl. Sci. 27 (1980) 1402.
\bibitem{Prigozhin} G. Prigozhin et al., ``The Depletion Depth of High Resistivity X-ray CCDs'', IEEE Trans. Nucl. Sci. 45 (1998) 903.
\bibitem{Janesick} J.R.~Janesick, ``Scientific Charge-Coupled Devices'', 
                   ISBN 9780819436986, SPIE Press Book (2001).
\bibitem{Tikkanen} T. Tikkanen, S. Kraft, F. Scholze, R.
Thornagel and G. Ulm, ``Characterising a Si(Li) detector element for the SIXA X-ray spectrometer'', Nucl. Instr. and Meth. A390 (1997) 329.
\bibitem{Hopkinson} G. R. Hopkinson et al., ``Proton Effects in Charge-Coupled Devices'', IEEE Trans. Nucl. Sci. 43 (1996) 614.
\bibitem{Hopkins94} I. H. Hopkins et al., ``Proton-Induced Charge Transfer Degradation in CCDs for 
Near-Room Temparature Applications'', IEEE Trans. Nucl. Sci. 41 (1994) 1984.
\bibitem{SR} W. Shockley and W. T. Read, ``Statistics of the Recombination of Holes and Electrons'', Phys. Rev. 87 (1952) 835.
\bibitem{Hall} R. N. Hall, ``Electron-Hole Recombination in Germanium'', Phys. Rev. 87 (1952) 387.
\bibitem{Kono} K. Kono et al., ``Interstitial carbon reactions in n-Si induced by high-energy 
proton irradiation'', Physica B 308 (2001) 265.
\bibitem{Albert} Albert J. P. Theuwissen, ``Solid-State Imaging with Charge-Coupled Devices'', Kluwer Academic Publishers (1995).
\end{thebibliography}
\end{document}